\documentclass[pre,aps,twocolumn,showpacs]{revtex4}

\usepackage{graphicx}
\usepackage{amsmath}
\usepackage{amsfonts}
\usepackage{amssymb}

\newcommand{\beqs}{\begin{eqnarray*}}
\newcommand{\eeqs}{\end{eqnarray*}}
\newcommand{\beq}{\begin{eqnarray}}
\newcommand{\eeq}{\end{eqnarray}}
\newcommand{\bit}{\begin{itemize}}
\newcommand{\eit}{\end{itemize}}
\newcommand{\eps}{\epsilon}
\newcommand{\np}{n^+}
\newcommand{\nm}{n^-}
\newcommand{\p}{\partial}
\newcommand{\dnp}{\delta n^+}
\newcommand{\dnm}{\delta n^-}
\newcommand{\jp}{j^+}
\newcommand{\jm}{j^-}

\newcommand{\al}{\alpha}

\newcommand{\dm}{\lambda_m}

\newcommand{\kp}{\mathbf{k}_\perp}
\newcommand{\kpm}{k_\perp}
\newcommand{\rp}{\mathbf{r}_\perp}
\newcommand{\vp}{\mathbf{v}_\perp}
\newcommand{\fp}{\mathbf{f}_\perp}
\newcommand{\epsm}{\epsilon_m}

\begin{document}
\title {Effective zero-thickness model for a conductive membrane driven by an electric field}
\author{Falko Ziebert$^1$}
\author{Martin Z. Bazant$^2$}
\author{David Lacoste$^1$}
\affiliation{$^1$ Laboratoire de Physico-Chimie Th\'eorique - UMR CNRS Gulliver 7083,
ESPCI, 10 rue Vauquelin, F-75231 Paris, France}
\affiliation{$^2$ Department of Chemical Engineering and Department of Mathematics,
Massachusetts Institute of Technology, Cambridge, MA 02139, USA}
\date{\today}
\begin{abstract}
The behavior of a conductive membrane in a static (DC) electric field is investigated theoretically.
An effective zero-thickness model is constructed based on a Robin-type boundary condition
for the electric potential at the membrane, originally developed for electrochemical systems.
Within such a framework, corrections to the elastic moduli of the membrane are obtained, which
arise from charge accumulation in the Debye layers
due to capacitive effects and electric currents through the membrane
and can lead to an undulation instability of the membrane.
The fluid flow surrounding
the membrane is also calculated, which clarifies issues regarding
these flows sharing many similarities with
flows produced by induced charge electro-osmosis (ICEO).
Non-equilibrium steady states of the membrane and of the fluid can be effectively described
by this method. It is both simpler, due to the
zero thickness approximation which
is widely used in the literature on fluid membranes,
and more general than previous approaches.
The predictions of this model are compared to
recent experiments on supported membranes in an electric field.
\end{abstract}
\pacs{87.16.-b, 82.39.Wj, 05.70.Np}

\maketitle

%%%%%%%%%%%%%%%%%%%%%%%%%%%%%%%%%%%%%%%%%%%%%%%%%%%%%%%%%%%%%%%%%%%%%%%%%%%%%%%
\section{Introduction}
\label{Intro}
%%%%%%%%%%%%%%%%%%%%%%%%%%%%%%%%%%%%%%%%%%%%%%%%%%%%%%%%%%%%%%%%%%%%%%%%%%%%%%%

Bilayer membranes formed from phospholipid molecules are
an essential component of the membranes of cells. The mechanical properties
of equilibrium membranes are characterized by two elastic moduli,
the surface tension and the curvature modulus \cite{seifert_mb_review:1997}, which typically
depend on the electrostatic properties of the membranes \cite{andelman}.
Understanding how these properties are modified when the membrane is driven
out of equilibrium is a problem of considerable importance to the
physics of living cells.
A membrane can be driven out of equilibrium in many ways, for instance
by ion concentration gradients or electric fields, either applied externally or generated internally.

The external application of electric fields on lipid films is used to
produce artificial vesicles
(by electroformation), as well as to create holes in the
membrane (by electroporation) \cite{electroporation:1989}.
Both processes are important for biotechnological
applications, they are widely used experimentally
although they are still rather poorly understood.
The generation of ion concentration gradients by internal means is controlled
in biological cells by membrane-bound ion pumps and channels,
which play key roles in many areas of biology
 \cite{hille}.

The non-equilibrium fluctuations of membranes including ion channels and pumps
were first analyzed in Refs.~\cite{PB,RTP} by means of an hydrodynamic theory.
Artificially made active membranes inspired by these ideas
were then studied experimentally \cite{jb_PRL,jb_PRE,Faris2009}.
Several theoretical studies followed, mainly motivated by the question of how to
model non-equilibrium effects produced by protein conformation changes \cite{gov,gautam,chen,lomholt}.
One limitation of existing active membrane models is that they do not
describe electrostatic effects associated with ion transport in details.
In previous papers by our group \cite{lacosteEL,lacosteEPJE},
we have addressed this limitation by studying a theoretical model for
a membrane with a finite conductivity transverse to the membrane plane
(due for instance to ion channels or pumps)
using electrokinetic equations \cite{hunter,Armand:2000,bazant:2004}.
Our work complements Ref.~\cite{lomholt_elect},
where the correction to the
elastic moduli of a membrane in a DC electric field
were calculated using an approach purely based on electrostatics (no currents).
It is also inspired by Ref.~\cite{kumaran} and Ref.~\cite{leonetti:2004}
where similar problems
were considered using electrokinetic equations.
In contrast to these studies, our approach focuses on the non-equilibrium case, where
electrokinetic corrections to the elastic moduli arise due to currents through the membrane.

In particular, a negative correction to the surface tension arises
due to capacitive effects, also called Lippman tension
\cite{sachs}. This negative tension leads to instabilities as can
be understood from the high-salt limit
\cite{pierre}. A first experimental proof of the destabilizing effect of the electric
field on a stack of lipid membranes was brought by X-ray scattering
studies \cite{salditt:05}. Recently,
the lowering of the tension due to electrostatic or electrokinetic effects
has been observed experimentally with supported membranes
subjected to an AC electric field \cite{charitat_EPJE} and
in active membranes \cite{Faris2009}.

The resulting flow fields around the undulating membrane are interpreted
within the framework of 'induced charge electro-osmosis' (ICEO)
\cite{murtsovkin,bazant:2004,squires2004}.
Similar flow patterns within vesicles subject to AC electric fields have been observed
experimentally and analyzed theoretically in Ref.~\cite{lipowsky:08}.
The deformation of lipid vesicles in alternating fields in various medium conditions
has been modeled theoretically in Refs.~\cite{petia:2009,svetina:2007}.
All these studies show that lipid membranes in electric fields present
a rich panel of possible behaviors \cite{dimova,lipowsky:08}.

This paper extends previous work \cite{lacosteEL,lacosteEPJE}, by
providing an effective zero-thickness membrane model that contains
both capacitive effects and ionic currents. In a first attempt \cite{lacosteEL},
a zero-thickness membrane model has been proposed
with the boundary condition (BC) of zero electric field at the membrane.
Although the shape of the potential was acceptable, the charge
distribution had the wrong sign and the elastic moduli were orders
of magnitude too small.
A model with finite membrane thickness and dielectric constant
has thus been considered in Ref. \cite{lacosteEPJE}, leading to correct
signs of the charge distribution and
correct orders of magnitude of the elastic moduli. However, this
model needed approximations and
finally numerical evaluations. In view of this, we present here
an improved zero-thickness model, by using the more realistic
BC of a dielectric interface sustaining Faradaic
currents \cite{bazant2005}. Although this Robin-type
BC has been introduced in Ref.
\cite{lacosteEPJE}, its consequences were not developed.
In particular this model leads to simple analytical expressions for the corrections
to the elastic constants of the membrane.
The model clearly captures both non-equilibrium effects due to ion currents and equilibrium
effects, of capacitive nature. We also calculate the flow field around the membrane,
which has in fact the opposite sign as compared to the one of Ref.
\cite{lacosteEPJE} for the zero-thickness case, and is thus
similar to standard ICEO flow fields.
The presented effective zero-thickness model for a DC-field driven
conductive membrane
is simple enough to be the starting point of more refined
further studies.

The work is organized as follows: in section \ref{Model} we describe the equations for the
charges in the electrolyte. A special emphasis is put on the boundary conditions
which is the crucial point here.
Then the base state solution corresponding to a flat membrane
is calculated in section \ref{base}.
In section \ref{O_hgen} we calculate the leading order
contributions to the electric and ion density fields
for a spatially modulated membrane height and
analyze the corresponding hydrodynamic flows around the membrane.
Using the boundary conditions for the
stress tensor at the membrane (which includes Maxwell and hydrodynamic stresses),
we calculate in section \ref{membr_fluct}
the growth rate of membrane fluctuations.
In section \ref{Disc}, our results are discussed and compared to
previous calculations and to related experiments.

%%%%%%%%%%%%%%%%%%%%%%%%%%%%%%%%%%%%%%%%%%%%%%%%%%%%%%%%%%%%%%%%%%%%%%%%%%%%%%%
\section{Model equations}
\label{Model}
%%%%%%%%%%%%%%%%%%%%%%%%%%%%%%%%%%%%%%%%%%%%%%%%%%%%%%%%%%%%%%%%%%%%%%%%%%%%%%%
%
We consider a steady (DC) current driven by a voltage $V$ between two electrodes at a fixed distance $L$,
applied to an initially flat membrane located initially at $z=0$.
The membrane is embedded in an electrolyte of monovalent ions with densities $n^+$
and $n^-$. The membrane has channels for both ion species but is itself neutral
(no fixed charges at the membrane).
The channels/pumps are assumed to be homogeneously distributed in the membrane
and enter only in the effective conductance $G$, as introduced below.
For the effect of non-uniform distributions of channels/pumps in membranes
we refer to Refs.~\cite{gautam,misbah}.
A point in the membrane is characterized by its Monge representation (valid in
the limit of small undulations) by introducing a height function $h(\rp)$,
where $\rp$ is a two-dimensional in-plane vector.

In the electrolyte, the governing equation for the electric potential $\phi$ is Poisson's equation
\beq
\label{Poisson_n}
\p_z^2\phi=-\frac{1}{\epsilon}\left(en^+ - en^-\right)\,.
\eeq
Here $e$ is the elementary charge
and $\epsilon$ is the dielectric constant of the electrolyte.
For the sake of simplicity, we assume a symmetric $1:1$ electrolyte, so that
far away from the membrane $n^+=n^-=n^*$.
We also assume that
the total system is electrically neutral.

The densities of the ion species are assumed to obey the
Poisson-Nernst-Planck equations for a dilute solution
\beq
\p_t n^\pm+\p_z J^{\pm}=0\,,
\eeq
with ionic current densities
\beq
J^{\pm}=D\left(-\p_z n^{\pm}\mp n^{\pm}\frac{e}{k_B T}\p_z\phi\right)\,,
\eeq
where $k_B T$ is the thermal energy. We have assumed that both ion types
have the same diffusion coefficient $D$, and neglected various corrections for
concentrated solutions~\cite{bazantACIS}.
We consider a steady state situation and use
the Debye-H\"uckel approximation by linearizing the
concentrations $n^\pm=n^*+\delta n^\pm$,
leading to
\beq
&&\p_z^2\phi=-\frac{e}{\epsilon}\left(\delta n^+ - \delta n^-\right)\,,\\
&&\p_z \left(-\p_z\delta\np-\frac{en^*}{k_B T}\p_z\phi\right)=0\,,\\
&&\p_z \left(-\p_z\delta\nm+\frac{en^*}{k_B T}\p_z\phi\right)=0\,.
\eeq

For symmetric binary electrolytes, it is useful to introduce {\it half} of the charge density~\cite{bazant2005,bazant2004PRE},
\beq\label{rhodef}
\rho=e\frac{\dnp-\dnm}{2}\,,
\eeq
as well as  the average concentration of the two ionic species,
\beq
c=e\frac{\dnp+\dnm}{2}\,.
\eeq
The latter quantity turns out not to be a relevant variable in the following since
it is decoupled from the field at small applied voltages~\cite{bazant2004PRE}.
Moreover, since we have considered a steady state
and a symmetric situation, there is no net particle current.
We also define
\beq
j^\rho&=&\frac{\jp-\jm}{2}=-D\p_z\left(\rho+\frac{e^2n^*}{k_B T}\phi\right)\,,
\eeq
which represents half of the electric current density,
and we arrive at the equations
\beq
\label{Poissondim}
&&\p_z^2\phi=-\frac{2}{\epsilon}\rho\,,\\
\label{PNPdim}
&&\p_z^2\rho+\frac{e^2n^*}{k_B T}\p_z^2\phi=0\,.
\eeq
Insertion of Eq.~(\ref{Poissondim}) into Eq.~(\ref{PNPdim}) leads to
\beq\label{rhokappaeq}
\left(\p_z^2-\kappa^2\right)\rho=0\,,
\eeq
where
\beq\label{kappa2def}
\kappa^2=\frac{2e^2n^*}{\epsilon k_B T}
\eeq
and $\kappa^{-1}=\lambda_D$ is the Debye length that defines the characteristic length scale for charge relaxation
in the electrolyte.

%%%%%%%%%%%%%%%%%%%%%%%%%%%%%%%%%%%%%%%%%%%%%%%%%%%%%%%%%%%%%%%%%%%%%%%%%%%%%%%
\subsection{Boundary conditions}
\label{BC}
%%%%%%%%%%%%%%%%%%%%%%%%%%%%%%%%%%%%%%%%%%%%%%%%%%%%%%%%%%%%%%%%%%%%%%%%%%%%%%%

At the electrodes located at $z=\pm \frac{L}{2}$, we externally impose the voltage leading to
\beq
\phi\left(z=\pm \frac{L}{2}\right)=\pm \frac{V}{2}\,.
\eeq
This BC is oversimplified for real electrodes, since it neglects interfacial
polarization across the double layers passing Faradaic currents \cite{bazant2005,biesheuvel2009galvanic},
which makes the voltage imposed across the electrolyte, outside the double layers,
different from the applied voltage. Since we focus on the membrane dynamics, however,
electrode polarization is inconsequential, and the voltage $V$ in the model simply serves
as a means to apply a steady DC current, which could be directly measured or imposed in experiments
testing our theory. We assume in the following that the distance between the electrodes is much larger
then the Debye length, $L\gg\lambda_D=\kappa^{-1}$.
In that case, the bulk electrolyte is quasi-neutral, $n^+=n^-=n^*$, with negligible charge density
(compared to the total salt concentration),
\beq
\rho\left(z=\pm \frac{L}{2}\right)
=0\,.
\eeq
Since the conductivity of a quasi-neutral electrolyte is constant,
the applied uniform current is equivalent to an applied electric field far from the membrane.

As we will see, the BC at the membrane is crucial to recover the
correct physical behavior.
In the simple zero-thickness model proposed in Ref.~\cite{lacosteEL}
the Neumann BC
\beq\label{BC_simple}
\p_z\phi_{|z=0}=0,
\eeq
was used for the potential, corresponding to
a vanishing electric field at the membrane.
Thus the dielectric mismatch between the electrolyte and the membrane
was accounted for only approximatively.
When compared to the full finite thickness
calculation, the agreement was poor.
To address this issue, a more general Robin-type BC was introduced \cite{lacosteEPJE}
\beq\label{RobinBC}
\dm\p_z\phi_{|z=0+}=\dm\p_z\phi_{|z=0-}=\phi(0^+)-\phi(0^-)\,,
\eeq
where
\beq\label{dmdef}
\dm=\frac{\epsilon}{\epsilon_m}d
\eeq
is a length scale that contains the membrane thickness $d$ and the ratio
of the dielectric constant of the electrolyte, $\epsilon$, and of the membrane, $\epsilon_m$.
This BC (with one side held at constant potential)
was originally developed for electrodes sustaining Faradaic
current~\cite{bazant2005,biesheuvel2009galvanic,itskovich1977,bonnefont2001}
or charging capacitively~\cite{bazant2004PRE,bazant:2004}. In that context the analog of our membrane
is a Stern monolayer of solvent molecules or a thin dielectric coating, such as a native oxide,
on a metallic surface, and $\lambda_m$ is denoted $\lambda_S$. Note that this BC has also been used
in Ref.~\cite{leonetti:2004}.

The modified boundary condition (\ref{RobinBC}) introduces a new dimensionless parameter~\cite{bazant2005},
\begin{equation}
\delta_m = \kappa\dm=\frac{\dm}{\lambda_D} = \frac{\epsilon\kappa}{\epsilon_m/d}=\frac{C_D}{C_m}\,.
\end{equation}
%which plays a central role in our analysis below.
For a blocking or ``ideally polarizable'' surface, which does not pass normal current
and only allows capacitive charging of the double layer, this parameter controls
the relative importance of the capacitance of the  surface (here, the membrane)
$C_m=\epsilon_m/d$ compared to that of the diffuse part of of the double layer,  $C_D=\epsilon\kappa$.
The BC (\ref{RobinBC}) then implies that these capacitances are effectively
coupled in series in an equivalent-circuit representation of the double layer~\cite{bazant2004PRE}.
For a surface sustaining normal current, either by electron-transfer reactions at an electrode
or by ionic flux through a membrane, the situation is more complicated.
It can be shown that the same electrostatic BC (\ref{RobinBC})
remains valid for a thin dielectric layer, as long as it has zero total free charge~\cite{bazantACIS},
which is typical for membranes containing a high density of fixed counter charge.
However, the same parameter $\delta_m$ no longer plays the role of a capacitance ratio.
Instead, it  controls the effect of diffuse charge on the normal current,
the so-called ``Frumkin correction'' to reaction kinetics in electrochemistry,
reviewed in Ref.~\cite{biesheuvel2009galvanic}. Two distinct regimes were first
identified in Ref.~\cite{bazant2005} in the context of electrolytic cells and
recently extended to galvanic cells~\cite{biesheuvel2009galvanic}:
(i) the ``Helmholtz limit" $\delta_m \gg 1$, where most of the double layer voltage
is dropped across the surface or membrane and the diffuse-layer has no effect
on the current, and (ii) the ``Gouy-Chapman limit" $\delta_m \ll 1$ where the diffuse layer
carries all of the voltage and thus determines the current.
In the Helmholtz limit, the Robin BC (\ref{RobinBC}) reduces to the Neumann BC (\ref{BC_simple})
used in Ref.~\cite{lacosteEPJE}, so that paper analyzed the limit where the diffuse charge
is small and has little effect on the current.
In this paper we consider the general case of finite $\delta_m$.

%%%%%%%%%%%%%%%%%%%%%%%%%%%%%%%%%%%%%%%%%%%%%%%%%%%%%%%%%%%%%%%%%%%%%%%%%%%%%%%
\section{Base state}
\label{base}
%%%%%%%%%%%%%%%%%%%%%%%%%%%%%%%%%%%%%%%%%%%%%%%%%%%%%%%%%%%%%%%%%%%%%%%%%%%%%%%

The base state of the problem is a flat membrane.
The electric field, assumed to be perfectly aligned in $z$-direction,
is then perpendicular to the membrane.
In the bulk fluid, the system is completely characterized
by the electrostatic potential
$\phi_0(z)$ (or by the field $E_0(z)=E^z_0(z)=-\p_z\phi_0$),
by the steady-state ion distribution $\rho_0(z)$,
as well as by the %(osmotic)
pressure $P_0(z)$. Inside the membrane, an internal electrostatic potential
$\phi^m_0(z)$ (and field $E_0^m$) is present.

%
%%%%%%%%%%%%%%%%%%%%%%%%%%%%%%%
%\subsection{Electrostatic potential and charge distribution}
%\label{elstat1}
%%%%%%%%%%%%%%%%%%%%%%%%%%%%%%%

Equation (\ref{rhokappaeq}) is readily solved leading to the charge distribution
\beq\label{charge}
\rho_0(z)=\left\{\begin{array}{cc}%{cc}
\rho_m e^{-\kappa z}\,\,&;\,\,z>0\\
-\rho_m e^{\kappa z}\,\,&;\,\,z<0
\end{array}\right.\,.
\eeq
Insertion into Eq.~(\ref{PNPdim}) and integrating once yields
the potential
\beq\label{field}
\phi_0(z)=\left\{\begin{array}{cc}%{cc}
\frac{2}{\epsilon\kappa^2}\left[\frac{j_m}{D}\left(z-\frac{L}{2}\right)-\rho_me^{-\kappa z}\right]
+\frac{V}{2}\hspace{-1mm}&;\,\,z>0\\
\frac{2}{\epsilon\kappa^2}\left[\frac{j_m}{D}\left(z+\frac{L}{2}\right)+\rho_me^{\kappa z}\right]
-\frac{V}{2}\hspace{-1mm}&;\,\,z<0
\end{array}\right.\hspace{-1mm}.\quad
\eeq
Here $j_m=-j^\rho$ is the electric current density and
\beq
\rho_m=\frac{\rho(z=0^+)-\rho(z=0^-)}{2}=:\frac{1}{2}[\rho_0]_{z=0}
\eeq
represents the jump in the charge density across the membrane.
We have introduced the notation
\beq
[f]_{z=a}=f(z=a^+)-f(z=a^-)\,,
\eeq
by which we denote the jump of
the field $f$ at position $z=a$.
Note that the jump in the charge density $\rho_m$ at the membrane can be
interpreted in terms of a surface dipole localized on the membrane. The existence of
this surface dipole is the physical reason for the discontinuity of the potential
at the membrane.

At the membrane, the Robin-type BC reads
\beq\label{RobinBC_dim}
\dm\p_z\phi_{|z=0+}=\dm\p_z\phi_{|z=0-}=[\phi]_{z=0}\,.
\eeq
Using Eq. (\ref{field}), the potential jump at the membrane is
\beq
[\phi_0]_{z=0}
=\frac{2}{\epsilon\kappa^2}\left(-\frac{j_m L}{D}-2\rho_m\right)+V.
\eeq

Although the membrane has a zero thickness in this model,
one can still define an internal field $E_0^m$ and an internal
potential $\phi_0^m(z)$. This is done by keeping a finite thickness $d$ at first,
and then take the limit $d \rightarrow 0$ \cite{lacosteEPJE}.
The continuity of the potential at the membrane boundaries then
implies %$[\phi_0]_{z=0}=[\phi_0^m]_{z=0}$. This corresponds to 
a constant internal field $E_0^m=-[\phi_0]_{z=0}/d$, where the
internal potential is $\phi^m_0(z)=[\phi_0]_{z=0}z/d$, or explicitly
\beq \label{internal potential}
\phi^m_0(z)=\frac{1}{d}\left[\frac{2}{\eps\kappa^2}\left(-\frac{j_m L}{D}-2\rho_m\right)+V\right]z.
\eeq

Using Eq. (\ref{RobinBC_dim}), one obtains the jump in the charge density
\beq\label{a_formula}
\rho_m=\frac{\frac{\epsilon\kappa^2}{2}V-\frac{j_m}{D}(L+\dm)}{2+\kappa\dm}\,.
\eeq
Two remarks on this derivation are in order: first, Eq.~(\ref{a_formula})
illustrates that the asymmetry of the charge distribution results either from the
accumulation of charges due to the applied voltage (a capacitive effect proportional to $V$, present also
for non-conductive membranes)
or due to ionic currents across the membrane
(a non-equilibrium effect proportional to $j_m$ present only for
conductive membranes).
Second, the expressions for $\rho(z)$, $\phi(z)$, $\rho_m$
derived above are independent of the response of the ion channels and thus remain {\it unchanged}
if a nonlinear ion channel response is used. Only the expression for the current $j_m$,
that enters as a parameter, will be affected by such a nonlinearity.
For precisely that reason, the general form of the base state does not depend on
the mechanism that has created the current (external due to an applied field or internal due to pumps).

To determine the current density $j_m$ at the membrane position, we use
a linear response approach (for nonlinear ionic response, see for instance
Refs.~\cite{leonetti,leonetti:2004,hille})
\beq
j^\rho_{|z=0}=-\frac{G}{e}[\mu^\rho]_{z=0}\,,
\eeq
where $\mu^\rho$ is the chemical potential per particle
and $G$
is the membrane conductivity per unit surface (across the membrane, not in-plane).
We assumed equal $G$ for both ion species.
In the bulk one has
$j^\rho=-\frac{eDn^*}{k_B T}\p_z\mu^\rho$,
which leads to the usual expression for the (electro-)chemical potential,
$\mu^\rho=k_B T\frac{\rho}{e n^*}+e\phi$.
By equating
\beq
-j_m=j^\rho_{|z=0}=-\frac{G}{e}\left(\frac{k_B T}{en^*}[\rho_0]_{z=0}+e[\phi_0]_{z=0}\right)\,,
\eeq
we finally arrive at the simple expression
\beq\label{sigma_form}
j_m=-j^\rho=\frac{GV}{1+\frac{2}{\epsilon\kappa^2 D}GL}\,.
\eeq
This relation is consistent with the usual electric circuit representation
of ion channels in a membrane that is surrounded by an electrolyte \cite{hille,lacosteEL}.
As far as the sign convention of the currents is concerned,
the cathode (towards where the cations drift) is located at $z=-L/2$ and the anode at $z=L/2$.
Thus $j_m$ is positive, in accordance with the usual convention
for transport of positive charges from the anode to the cathode.
%so that $j^\rho$ is a current in $-z$-direction.
Insertion of $j_m$ into $\rho_m$ yields
\beq\label{a_form}
\rho_m=\frac{\epsilon\kappa^2}{2}V
\frac{1-\frac{2}{\epsilon\kappa^2 D}G\dm}{\left(1+\frac{2}{\epsilon\kappa^2 D}GL\right)(2+\kappa\dm)}\,.
\eeq
The jump in the charge at the membrane is thus positive,
$\rho_m>0$, when $\lambda_m<\epsilon\kappa^2 D/(2G)$.
For reasonable parameters for a biological membrane (see section \ref{Disc}),
$\rho_m>0$ holds. A negative charge jump is only possible if the electrolyte
contains very low salt
and in addition the membrane is strongly conductive.
This is the reason why the zero thickness model with the simple BC, Eq.~(\ref{BC_simple}),
is not realistic for biological membranes,
since it corresponds to the case $\dm=\infty$, implying $\rho_m<0$.
We note that by performing the limit $\dm\rightarrow\infty$ in Eq.~(\ref{a_form}),
one regains the results   for the zero-thickness model of Ref. \cite{lacosteEPJE} with BC
Eq.~(\ref{BC_simple}).

%
%
%%%%%%%%%%%%%%%%%%%%%%%%%%%%%%%
%\subsection{Maxwell stresses}
%\label{membr1}
%%%%%%%%%%%%%%%%%%%%%%%%%%%%%%%

To complete the description of the base state,
we have to consider the total stress tensor
\beq\label{stresstensor}
\tau_{ij}=-P\delta_{ij}+\eta\left(\p_i v_j+\p_j v_i\right)
+\epsilon\left(\hspace{-1mm}E_i E_j -\frac{1}{2}\delta_{ij}E^2\hspace{-1mm}\right)\,\,\,
\eeq
at the membrane.
It contains the pressure,
the viscous stresses in the fluid
and the Maxwell stress due to the electrostatic field. We denote by $\eta$ the viscosity
of the electrolyte and by $\mathbf{v}$ its velocity field. The electric field is given by
$\mathbf{E}=-\nabla\phi$.

In the base state, where the membrane is flat and the electric field is oriented in $z$-direction,
from $\nabla\cdot\tau=0$ we get
$\p_z P_0=\frac{\epsilon}{2}\p_z\left(\left(\p_z\phi_0\right)^2\right)=-2\rho_0\p_z\phi_0$.
By using Eqs. (\ref{charge}, \ref{field}) and imposing $P(z\rightarrow\infty)=0$, this is readily solved
leading to
\beq
\label{pressure profile base state}
P_0(z>0)=\frac{4}{\epsilon\kappa^3}\left(\frac{\rho_m j_m}{D}e^{-\kappa z}
+\frac{\kappa \rho_m^2}{2}e^{-2\kappa z}\right)\,,
\eeq
and similarly with $z\rightarrow -z$ for $z<0$.
For the stress we thus get
\beq
\tau_{zz,0}(z>0)=\tau_{zz,0}(z<0)=\frac{2}{\epsilon\kappa^4 D^2}j_m^2\,.
\eeq
Note that the stress is constant and is due to the current density $j_m$, with no contributions
from the induced charges $\rho_m$. At the membrane, the stress is balanced.
In addition to the part of the stress tensor due to the electric field in the electrolyte,
there also is the part due to the electric field inside the membrane, already mentioned above.
For the force balance in the base state however, this contribution vanishes is thus not important.

%%%%%%%%%%%%%%%%%%%%%%%%%%%%%%%
\section{Leading order contribution of membrane fluctuations}
\label{O_hgen}
%%%%%%%%%%%%%%%%%%%%%%%%%%%%%%%

In the following, we derive the corrections to the base state to first order in the
membrane height $h(\rp)$. From such a calculation, we obtain the growth
rate of membrane fluctuations
by imposing the BC for the normal stress
at the membrane. From this growth rate, we then can identify electrostatic and electrokinetic corrections to the
elastic moduli of the membrane.

%%%%%%%%%%%%%%%%%%%%%%%%%%%%%%%
\subsection{Electrostatics to first order in $h(\rp)$}
\label{O_h}
%%%%%%%%%%%%%%%%%%%%%%%%%%%%%%%

We use here the quasi-static approach \cite{lomholt_elect,lacosteEPJE}
by assuming
that membrane fluctuations
are much slower than the characteristic diffusion time
$\tau_D=\frac{1}{D\kappa^2}$ of the ions
to diffuse on a Debye length.
With the definition of the Fourier transform
$f(\kp,z)=\int d\rp e^{-i\kp\cdot\rp}f(\rp,z)$,
we expand the electric field and ion density as
\beq
\phi(\kp,z)&=&\phi_0(z)+\phi_1(\kp,z)\,,\\
\rho(\kp,z)&=&\rho_0(z)+\rho_1(\kp,z)\,,
\eeq
where $\kp$ lies in the plane defined by the membrane and $\phi_0(z)$, $\rho_0(z)$ are the base state
solutions given by Eqs. (\ref{charge}, \ref{field}).
The governing equations
to this order read
\beq
\label{cf1}&&\hspace{-3mm}\,\,\,\left(\p_z^2-\kpm^2\right)\phi_1(\kp,z)+\frac{2}{\epsilon}\rho_1(\kp,z)=0\,,\\
\label{cf2}&&\hspace{-3mm}\left(\p_z^2-\kpm^2\right)
\left(\rho_1(\kp,z)+\frac{\epsilon}{2}\kappa^2\phi_1(\kp,z)\right)=0\,,
\eeq
and for the particle currents at the membrane one has the condition
\beq
\label{cf3n}&&\hspace{-3mm}D\p_z\hspace{-1mm}\left(\rho_1+\frac{\epsilon}{2}\kappa^2
\phi_1\hspace{-1mm}\right)_{|z=h}
\hspace{-1mm}
=G\left(\frac{2}{\epsilon\kappa^2}\left[\rho_1\right]_{z=0}\hspace{-1mm}+[\phi_1]_{z=0}\right).\quad\quad
\eeq
Eqs. (\ref{cf2}, \ref{cf3n}) and the BCs at infinity
(since $L\gg\lambda_D$, we can use
$\phi_1(\kp,\pm \infty)=\rho_1(\kp,\pm \infty)=0$ far from the membrane)
are satisfied by choosing
\beq\label{rho1phi1}
\rho_1=-\frac{\epsilon}{2}\kappa^2\phi_1
\eeq
(and $n^+_1=-n^-_1$ implying $\rho_1=e n^+_1$).
It follows that to first order in the height,
one has a zero flux condition at the membrane.
Accordingly, the zeroth order solution enters in the equations
for the first order solution $\phi_1$ only via the boundary conditions.

It remains to solve the Poisson equation (\ref{cf1}),
\beq
\left(\p_z^2-\kpm^2-\kappa^2\right)\phi_1=0\,.
\eeq
We introduce
\beq\label{ldef}
l^2=\kpm^2+\kappa^2\,,
\eeq
with $l^{-1}$ as the characteristic length for the electrostatic problem near the
slightly undulated membrane,
and easily get $\phi_1=A^\mp e^{\mp lz}$
for $z>0$ and $z<0$ respectively. To determine the constants $A^\pm$,
we expand
\beq
\p_z\phi_{|z=h(\rp)}&=&\p_z\phi_{0|z=h}+\p_z\phi_{1|z=h}\nonumber\\
&=&\p_z\phi_{0|z=0}+\p_z^2\phi_{0|z=0}\cdot h(\rp)\nonumber\\
&&+\p_z\phi_{1|z=0}+O(h^2)\,,
\eeq
to prescribe the BC
at the membrane in first order.
Using this expansion in the Robin-type BC, Eq. (\ref{RobinBC_dim}),
we get
\beq
\label{phi1eq}\phi_1(\kp,z)&=&-\frac{2}{\epsilon}\rho_m \frac{h(\kp)}{l}e^{-lz}\,,\\
\rho_1(\kp,z)&=&
\kappa^2\rho_m  \frac{h(\kp)}{l}e^{-lz}\,,
\eeq
for $z>0$ and a symmetric expression (with $e^{lz}$) for $z<0$.

%%%%%%%%%%%%%%%%%%%%%%%%%%%%%%%
\subsection{Linear hydrodynamic flow}
\label{hydro}
%%%%%%%%%%%%%%%%%%%%%%%%%%%%%%%

When the membrane starts to undulate with small amplitude $h(\rp)$, a flow is induced
in the surrounding electrolyte.
Using again the quasi-static assumption and low Reynolds number,
this flow is governed by incompressibility and the Stokes equation,
\beq\label{incomp}
\nabla\cdot \mathbf{v}&=&0\,,\\
-\nabla p+\eta\Delta \mathbf{v}+\mathbf{f}&=&0\,,
\eeq
where $\mathbf{f}$ is a body force density due to the electric field.
Introducing the triad \cite{seifert_mb_review:1997,bickel} of unit vectors
$(\hat{\mathbf{k}}_\perp,\hat{\mathbf{n}},\hat{\mathbf{t}})$
with $\hat{\mathbf{k}}_\perp=\kp/\kpm$, $\hat{\mathbf{n}}=\hat{z}$
and $\hat{\mathbf{t}}=\hat {\mathbf{k}}_\perp \times\hat{\mathbf{n}}$,
we get
\beq\label{inc1}
\p_z v_z+i\kp\cdot\vp&=&0\,,\\
\label{vp}-i\kp p+\eta\left(\p_z^2-\kpm^2\right) \vp +\fp &=&0\,,\\
\label{vz}-\p_z p+\eta\left(\p_z^2-\kpm^2\right) v_z +f_z &=&0\,,\\
\label{vt}\eta\left(\p_z^2-\kpm^2\right) v_t +f_t &=&0\,.
\eeq

Two forces drive the flow: one is given by the coupling to the membrane,
enters via the BC, and is discussed below.
The second one is
the bulk force due to the electric field acting on the charge distribution and reads
$\mathbf{f}=Q\mathbf{E}=-2\rho\nabla\phi$.
Note that one has to use the total charge density, $Q=2\rho$.
To leading order in the membrane height, this driving force is
$\mathbf{f}=-2\rho_0\nabla\phi_1-2\rho_1\nabla\phi_0+O(h^2)$
with components
\beq
\label{fp}\fp&=&-2\rho_0(z)i\kp\phi_1(\kp,z)\,,\\
\label{fz}f_z&=&-2\rho_0(z)\p_z\phi_1(\kp,z)-2\rho_1(\kp,z)\p_z\phi_0(z)\,,\,\quad
\eeq
and $f_t=0$. Because of the latter, Eq. (\ref{vt}) is decoupled and trivial.
The equations for $v_z$ and $\vp$ can be decoupled as follows:
Using incompressibility,
Eq. (\ref{inc1}), for
the perpendicular fluid velocity in Eq. (\ref{vp}),
one obtains for the pressure
\beq\label{press_dim}
p=-\eta\p_z v_z+\frac{\kp\cdot\fp}{i\kpm^2}+\frac{\eta}{\kpm^2}\p_z^3v_z\nonumber\,.
\eeq
Insertion into Eq. (\ref{vz}) yields a single equation for $v_z$.
For $z>0$ one has to solve
\beq
\left(\p_z^2-\kpm^2\right)^2 v_z
&=&\frac{2}{\eta}\kpm^2 \phi_1\p_z\left(\rho_0+\frac{\epsilon}{2}\kappa^2\phi_0\right)\nonumber\\
&=&- \frac{4}{\epsilon D \eta}\rho_m j_m\frac{\kpm^2 h(\kp)}{l}e^{-lz}  \,.
\eeq
Imposing $v_z(z\rightarrow\pm\infty)=0$, the solutions are of the form
\beq
v_z=\left(B^\mp +C^\mp z\right)e^{\mp\kpm z}+F^\mp e^{\mp lz}\,,
\eeq
for $z>0$ and $z<0$ respectively.
The coefficients $F^{\pm}$ are determined by the driving force $\mathbf{f}$,
but two more BCs are needed.
At the membrane, continuity of the normal velocity imposes
\beq
\label{vzcond}v_z(0^+)=v_z(0^-)&=&\p_t h(\rp)=sh(\kp),
\eeq
where we have introduced the growth rate for membrane fluctuations, $s$,
from the temporal Fourier representation
$h(t)\propto e^{st}$. Note that $s$ is also a function of $\kp$.

The continuity of the tangential velocity, $\vp(0^+)=\vp(0^-)=0$,
together with the incompressibility implies a second BC for $v_z$, namely
\beq
\p_z v_{z|0^+}=\p_z v_{z|0^-}=0 \,.
\eeq
This allows to solve the complete linear hydrodynamics problem.
With the notation
\beq
\al=\frac{4\rho_m j_m}{\epsilon D \eta \kappa^4}\,,
\eeq
which quantifies the amplitude of the ICEO flow, the velocity and pressure fields
in the domain $z>0$ read
\begin{widetext}
\beq
\label{vzfull}
v_z(z>0)&=&h(\kp)\left[s\left(1+\kpm z\right)e^{-\kpm z}
-\al\kpm\left(\kpm z-\frac{\kpm}{l}-\frac{\kpm^2 z}{l}\right)e^{-\kpm z}
-\al\frac{\kpm^2}{ l}e^{-lz}\right]\,,\\
\label{vpfull}
\vp(z>0)&=&h(\kp)\,i\,\kp\left[  - s  z e^{-\kpm z}
- \al \left(1-\kpm z+\frac{\kpm^2 z}{l}   \right) e^{-\kpm z}
+ \al e^{-lz} \right]\,,\\
\label{pfull}
p(z>0)&=&h(\kp)\,\eta\left[
2\kpm\left(s-\al\left(\kpm-\frac{\kpm^2}{l}\right)\right)e^{-\kpm z}
+\al e^{-lz}
+\frac{4\rho_m^2}{\epsilon\eta l}e^{-(l+\kappa)z}
\right]\,.\quad\quad
\eeq
\end{widetext}
The solutions for $z<0$
can be obtained by symmetry operations:
$v_z(z<0)$ is obtained by performing the mirror operation with respect to the plane
defined by the membrane, $z\rightarrow-z$, in the formula for $v_z(z>0)$. Similarly
$-\vp(z<0)$ and $-p(z<0)$ are obtained by doing this operation on $\vp(z>0)$ and $p(z>0)$,
respectively.

In the absence of electric effects, $\rho_m=0=j_m$, one gets the typical flow
induced by a membrane bending mode
\cite{BrochardLennon,levine}.
An additional flow field due to the membrane currents arises that
has the form of an ICEO flow \cite{bazant:2004}.
A detailed discussion of this effect is postponed to section \ref{Disc},
but we stress that the additional flow is purely due to membrane conductivity
(since $\al\propto j_m\propto G$),
and is a non-equilibrium effect.
For non-conductive membranes
this flow vanishes but
there still is charge accumulation ($\rho_m\neq0$) in the Debye layers,
leading via the pressure field, Eq. (\ref{pfull}), to
corrections to surface tension and bending rigidity proportional to $\rho_m^2$.

%%%%%%%%%%%%%%%%%%%%%%%%%%%%%%%
\section{Growth rate of membrane fluctuations}
\label{membr_fluct}
%%%%%%%%%%%%%%%%%%%%%%%%%%%%%%%

To discuss the dynamics of the membrane, we still have to determine
the growth rate $s$ of the membrane.
The elastic properties of the membrane are described by the standard
Helfrich free energy
\beq \label{mb free energy1}
F_H=\frac{1}{2} \int d^2 \rp [ \Sigma_0 \left( \nabla h \right)^2 + K_0 \left( \nabla^2 h
\right)^2  ],
\eeq
where $\Sigma_0$ is the bare surface
tension
and $K_0$ the bare bending modulus of the membrane.
Force balance on the membrane implies that the restoring force
due to the membrane elasticity is equal to the discontinuity of the
normal-normal component of the stress tensor defined in Eq.~(\ref{stresstensor})
\beq
\label{BC_stress_normal}-\tau_{zz}(0^+) + \tau_{zz}(0^-)=-[\tau_{zz,1}]_{z=0}
= -\frac{\partial F_H}{\partial h(\rp)}\,.
\eeq
We should stress that the coupled electrostatics-hydrodynamics problem under investigation
can {\it not} be formulated only in terms of bulk forces, i.e. $\mathbf{f}$
and the divergence of a stress tensor,
because the hydrodynamic and Maxwell stress tensors enter the BC (\ref{BC_stress_normal}) explicitly.
For this reason, the force localized on the membrane surface is a priori unknown, i.e.
must be determined by BCs for the velocity and the stress.
Eq. (\ref{BC_stress_normal}) leads to
\beq\label{gen_disp_dim}
-[\tau_{zz,1}]_{z=0}
=\left(-\Sigma_0\kpm^2-K_0\kpm^4\right)h(\kp)\,,
\eeq
which determines the growth rate $s=s(\kp)$ entering the stresses.
The total normal stress at the membrane is
\beq\label{full_stress}
\tau_{zz,1}=\hspace{-1mm}
\left[-P+2\eta\p_z v_z+\frac{\epsilon}{2}\left(\p_z \phi\right)^2
-\frac{\epsilon_m}{2}\left(\p_z \phi^m\right)^2\right]_{|z=h}\,%\,,
\eeq
to linear order in $h$.
Note that we have included here the electrostatic contribution
stemming from the field inside the membrane (with potential $\phi^m$).
This contribution is particularly significant in the high salt limit,
where effects due to the Debye layers become
negligible. This inside contribution enters with opposite
sign than the outside contribution because
of the difference of orientation of the normal (see appendix \ref{Widom}
and Ref. \cite{lomholt_elect}).

The outer electrostatic contribution reads
\beq
\epsilon\left[(\p_z\phi_0)(\p_z\phi_1)\right]_{|z=0}
+h\p_z\left[-P_0+\frac{\epsilon}{2}\left(\p_z\phi_0\right)^2\right]_{|z=0}\,,\nonumber
\eeq
where the second contribution vanishes since the term in the bracket is a constant
(the stress is balanced, see above).
The electrostatic contribution from inside the membrane can be expressed analogously,
and the normal-normal stress difference at the membrane reads
\beq
[\tau_{zz,1}]_{z=0}\hspace{-1mm}&=&\hspace{-1mm}-[p]_{z=0} + 2\eta[\p_z v_z]_{z=0}\nonumber\\
& &\hspace{-1.7cm}+\epsilon\left[(\p_z\phi_0)(\p_z\phi_1)\right]_{z=0}
-\epsilon_m\left[(\p_z\phi^m_{0})(\p_z\phi^m_{1})\right]_{z=\pm d/2}\,\hspace{-1mm}.\quad\quad
\eeq
By means of Eq. (\ref{vzfull})
one easily verifies $[\p_z v_z]_{z=0}=0$ due to the symmetry given above.
For the pressure difference, Eq. (\ref{pfull}) implies $[p]_{z=0}=2p(0^+)$ and
after reexpressing $l$ by $\kpm$
and expanding in powers of $\kpm$, one obtains
\beq
[p]_{z=0}\hspace{-1mm}&=&\hspace{-1mm}h(\kp)\left[
8\left(\frac{\rho_m j_m}{\epsilon D \kappa^2}+\frac{\rho_m^2}{\epsilon\kappa}\right)
+4\eta s\kpm\nonumber\right.\\
&&\hspace{-1.4cm}+\left.4\hspace{-1mm}\left(-\frac{\rho_m^2}{\epsilon\kappa^3}
-\frac{4\rho_m j_m}{\epsilon D \kappa^4}\right)\kpm^2
+16\frac{\rho_m j_m}{\epsilon D \kappa^5} \kpm^3
+3\frac{\rho_m^2}{\epsilon\kappa^5} \kpm^4 \right]\hspace{-1mm}.\quad\quad
\eeq
The electrostatic contribution from the electrolyte reads
\beq
\epsilon\left[(\p_z\phi_0)(\p_z\phi_1)\right]_{z=0}=h(\kp)\,8
\left(\frac{\rho_m j_m}{\epsilon D \kappa^2}+\frac{\rho_m^2}{\epsilon\kappa}\right)\,,
\eeq
which exactly cancels the $\kpm$-independent contribution of $[p]_0$.
The calculation of the electrostatic contribution from inside the membrane is
slightly more involved and is detailed in Section \ref{ins_memb_new}.
The result is
\beq\label{contrib_ins}
\epsilon_m\left[(\p_z\phi^m_0)(\p_z\phi^m_1)\right]_{z=\pm d/2}&&\nonumber\\
&&\hspace{-4.2cm}=-h(\kp)\epsm (E^m_0)^2\hspace{-1mm}\left[d \kpm^2 +
 \left(-\frac{d^3}{12}+\frac{\rho_m}{E^m_0}\frac{d}{\eps\kappa^3}\right)
 \kpm^4\right]\hspace{-1mm},\quad\quad
\eeq
where $E^m_0$ is the zeroth order electric field inside the membrane introduced in
section \ref{base}.

By collecting all the contributions in Eq. (\ref{gen_disp_dim}),
the growth rate $s(\kpm)$ finally has the form
\beq\label{disp_fin}
\eta \kpm s(\kpm)&=&
-\frac{1}{4}\left(\Sigma_0+\Delta\Sigma\right)\kpm^2\nonumber\\
&&-\Gamma_\kappa\kpm^3
-\frac{1}{4}\left(K_0+\Delta K\right)\kpm^4\,.\quad
\eeq
The electrostatic corrections to the surface tension,
$\Delta\Sigma=\Delta\Sigma_\kappa+\Delta\Sigma_m$, and to
the bending modulus, $\Delta K=\Delta K_\kappa+\Delta K_m$
have been decomposed into outside contributions (due to the Debye layer, index $\kappa$)
and inside contributions (due to the voltage drop at the membrane, index $m$).
They are given by
\beq\label{deltasigma}
\Delta\Sigma_\kappa&=&-4\frac{\rho_m^2}{\epsilon\kappa^3}-16\frac{\rho_m j_m}{\epsilon\kappa^4 D}\,,\\
\label{deltak}
\Delta K_\kappa&=&\frac{3\rho_m^2}{\epsilon\kappa^5}
\eeq
for the contributions due to the Debye layers
and by
\beq\label{deltasm}
\Delta\Sigma_m&=&-\epsm (E^m_0)^2 d\,,\\
\label{deltakm}
\Delta K_m&=&\epsm (E^m_0)^2 \left(\frac{d^3}{12}-\frac{\rho_m}{E^m_0}\frac{d}{\eps\kappa^3}\right)
\eeq
for the contributions due to the field inside the membrane.
We note that the ratio $\Delta K /\Delta \Sigma$ is independent
of the applied voltage and scales with the square of the membrane
thickness.

An independent check of Eqs (\ref{deltasigma}, \ref{deltasm})
is provided by a direct integration of the lateral pressure profile as shown in
appendix \ref{Widom}. This route avoids the consideration of hydrodynamics but is limited
to the calculation of the surface tension correction.

In Eq.~(\ref{disp_fin}), we also obtain a purely non-equilibrium correction
\beq
\Gamma_\kappa=\frac{4\rho_m j_m}{\epsilon\kappa^5 D}=\frac{\eta}{\kappa}\alpha\,.
\eeq
It corresponds to a term proportional to $k_\perp^3$ in the effective
free energy of the membrane, which is forbidden for an equilibrium membrane but
allowed in non-equilibrium. This particular contribution
arises due to the electroosmotic flows around the membrane as can be shown
from a simple calculation using the Helmholtz-Smoluchowski equation for the
electro-osmotic slip velocity near the membrane \cite{lacosteEPJE}.

Two particular limits have been considered before: first, the case of a
non-conductive membrane ($j_m =0, \rho_m \neq 0$) with an arbitrary amount of salt,
where we recover results obtained recently by Ambj\"{o}rnsson
{\it et al.} \cite{lomholt_elect}, as further discussed in the
next section. Second, the high salt limit ($\kappa\rightarrow\infty$)
of a conductive membrane. Here, one finds that the correction to the surface tension is only
due to the inside field, since $\Delta\Sigma_\kappa\rightarrow 0$
and $\Delta\Sigma_m\rightarrow -\epsm \frac{V^2}{d}$, as has been calculated in
Ref. \cite{pierre}. The corrections to the bending modulus
were not considered in that reference, and are given from our calculation by
$\Delta K_\kappa\rightarrow0$ and $\Delta K_m\rightarrow \epsm
\frac{V^2}{12}d$, as in the non-conductive case.
In this limit, $\Delta K_m /\Delta \Sigma_m=-d^2/12$.

%%%%%%%%%%%%%%%%%%%%%%%%%%%%%%%%%%%%%%%%%%%%%%%%%%%%%%%%%%%%%%%%%%%%%%%%%%%%%%%
\section{Discussion}
\label{Disc}
%%%%%%%%%%%%%%%%%%%%%%%%%%%%%%%%%%%%%%%%%%%%%%%%%%%%%%%%%%%%%%%%%%%%%%%%%%%%%%%

\subsection{Applications of the model to experiments}

Recently, S. Lecuyer {\it et al.}
\cite{charitat_EPJE}
%and by L. Malaquin {\it et al.} by X-ray reflectivity. In both
have investigated a pair of nearby
membrane bilayers in
an electric field by neutron reflectivity. The first bilayer was
close to the bottom electrode and used to protect the second one from
interacting with the wall. Since the bare values of the elastic moduli %,
%before the application of the field,
were known from X-ray off-specular experiments
for a similar system \cite{charitat_PNAS}
($\Sigma_0 \simeq 0.5 $mNm$^{-1}$ and $K_0 \simeq 40 k_BT$),
the surface tension correction was extracted from the data
under the assumption that the bending modulus is not affected by the field.
%This is, according to our calculations, a strong approximation.
The experiments were performed in an
AC electric field at several frequencies.
For the lowest frequency ($10{\rm Hz}$),
the electrostatic correction to the surface tension was obtained to be $\Delta \Sigma \simeq -3$mNm$^{-1}$.

%i.e. of the same order as the bare surface tension $\Sigma_0$,
%and $\Delta K \simeq 185 \pm 15 {\rm k_B T}$ for the correction to
%the bending modulus, one order
%of magnitude higher than the bare modulus, $K_0$.
In the experimentally probed regime of low salt (D$_2$O was used as the electrolyte),
the electrostatic corrections to the elastic moduli depend rather sensitively
on both the amount of salt and on the ionic conductivity of the
membrane, as we will see below. Moreover, in the above experiment,
the correction to the bending modulus was not measured.
Thus we restrain ourselves to a comparison of orders of magnitude only.
The dielectric constants are  
$\epsilon=80\epsilon_0$ for the electrolyte %(water)and
$\epsilon_m=2\epsilon_0$ for the membrane.
The membrane thickness is typically $d=5{\rm nm}$ leading to $\dm=\frac{\epsilon}{\epsilon_m}d =200{\rm nm}$.
The diffusion coefficient of ions is of the order of $D=10^{-9}{\rm m}^{2}{\rm s}^{-1}$,
and the viscosity $\eta=10^{-3}{\rm Pa}\,{\rm s}$. % (water).
%In the experiments,
%milli-Q water was used ({\it still true?}), which corresponds to
%$\lambda_D=50-150{\rm nm}$.
The distance between the electrodes was about $L=1{\rm mm}$, while the
voltage was in the $1$-$5{\rm V}$ range. Assuming that the
membrane is non-conductive, $G=0$, for $\kappa=2\cdot10^7{\rm m}^{-1}$
and  $V=1{\rm V}$, our model yields
$\Delta\Sigma\simeq-2\cdot $mNm$^{-1}$ and $\Delta K\simeq190{\rm k_B T}$.
The model thus successfully accounts for the
order of magnitude of the electrostatic correction to the
surface tension. However, it also shows that the bending modulus increases
about five times. In order to obtain an
experimental test of the model, it would be interesting to measure the correction
to the surface tension and to the bending modulus
simultaneously. We would also like suggest
to carry out experiments in which the applied electric
field or the ionic strength would be varied. Another interesting
possibility would be to study membranes of different
conductivities or thicknesses in an applied electric field.

% Further work is also needed on the theory side. One limitation of
% the present model is the Debye-H\"uckel
% approximation, which only holds when the potential satisfies
% everywhere the condition $\frac{e\phi}{4k_B T}\ll1$ \cite{hunter}.
% We have found that the charge accumulation on the membrane
% is too large in the experiment discussed above with supported membranes
% for this approximation to hold, since one has
% $\frac{e\phi}{4k_B T} \simeq 2$. It would thus be interesting
% to solve the nonlinear Poisson-Nernst-Planck equations and
% otherwise proceed similarly as proposed in this work,
% in order to describe the behavior of membranes close to high charge densities.

A second field of application of the model are active membranes,
which are (artificial) lipid vesicles containing ionic pumps such as bacteriorhodopsin
\cite{jb_PRL,jb_PRE,Faris2009}. In these experiments, no external electric field is applied.
Instead the pumps are activated by light to transport protons across the membrane.
In Ref.~\cite{Faris2009}, a lowering of the membrane tension produced by the activity of the
pumps has been reported, which could be due to an accumulation
of charges near the membrane, as discussed here.
The specificity of that experiment is that this charge accumulation would result from
the activity of the pumps rather than from an applied electric field.
However, it is difficult to make a precise comparison between the
experiments and the present theory, because only the correction to the surface tension
is accurately measured
and many aspects of the transport of ions are unknown.
Nevertheless, if we assume that the passive state of that experiment corresponds 
to a non-conductive membrane ($G=0$) and the active state to a membrane with $G=10S$m$^{-2}$,
and if we use a typical transmembrane potential of the order of $50 {\rm mV}$,
we get the same order of magnitude for the observed tension lowering, $3 \cdot 10 ^{-7}$Nm$^{-1}$,
if we account for the rather high amount of salt with $\kappa \simeq 5 \cdot 10^8$m$^{-1}$.
We also find that there is no measurable difference for the bending modulus between the active
and passive state, as observed experimentally. The model predicts that a current density
of $j_m \simeq 1$Am$^{-2}$ arises when
the pumps are active, which corresponds to an overall current of $1$pA on a vesicle of size 1$\mu$m.
To better compare to the model, again it would be desirable
to have experiments in varying conditions (ionic strength and
conductivity of the membrane, for instance). Another interesting possibility would be
to measure the membrane current and the transmembrane
potential in the course of the experiment, for instance using patch-clamp techniques.

\subsection{Effect of salt and membrane conductivity}

\begin{figure}[t]
    \centering
    \includegraphics[width=0.45\textwidth]{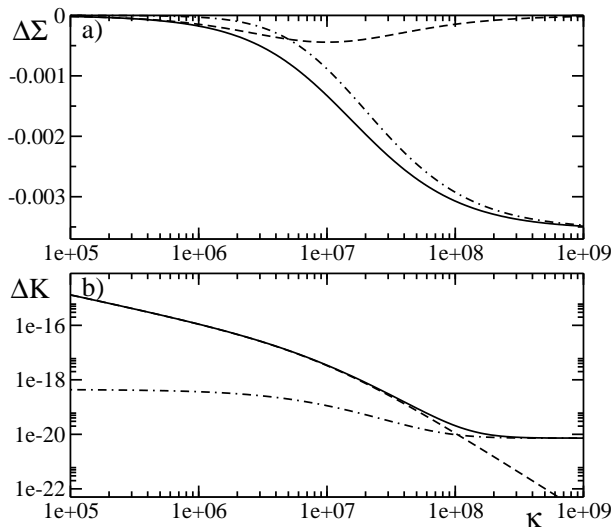}
    \caption{\label{DSdetailG0}
    Panel a) shows the electrostatic corrections
    to the surface tension (in units of N$\cdot$m$^{-1}$) 
    and panel b) those to the bending modulus (units $J$)
    as a function of $\kappa$ (units m$^{-1}$)
    in the non-conductive case, $G=0$.
    Dashed lines: contributions due to the Debye layer
    ($\Delta\Sigma_\kappa$ and $\Delta K_\kappa$ respectively).
    Dash-dotted lines: contributions due to the field inside the membrane
    ($\Delta\Sigma_m$ and $\Delta K_m$ respectively).
    Solid lines: sum of both corrections.
    The figure was made with the parameters given in the text and $V=1{\rm V}$,
    $L=1{\rm mm}$.
    }
\end{figure}

\begin{figure}[t]
    \centering
    \includegraphics[width=0.45\textwidth]{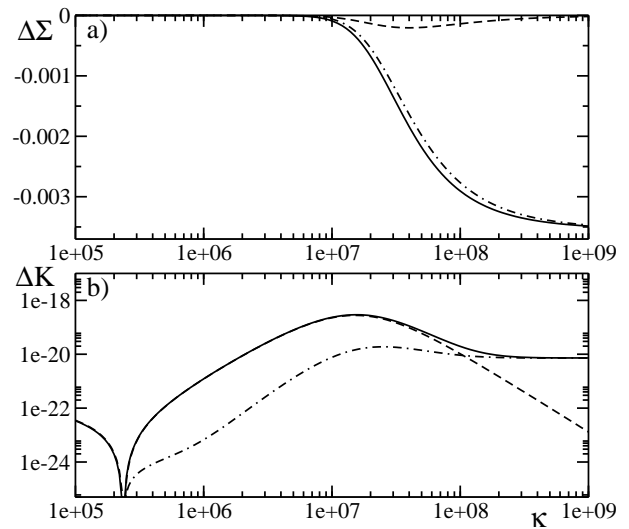}
    \caption{\label{DSdetailGfin}
    Electrostatic corrections
    to the surface tension (panel a) and to the bending modulus (panel b)
    as a function of $\kappa$ in a slightly conductive case ($G=0.1 S{\rm m}^{-2}$).
    Dashed lines: contributions due to the Debye layer.
    Dash-dotted lines: contributions due to the field inside the membrane.
    Solid line: sum of both corrections. Parameters as in previous figure except
    for $G$.
    }
\end{figure}

With the expressions for the jump of the charge density at the
membrane, Eq. (\ref{a_form}), and for the current density, Eq. (\ref{sigma_form}),
we can discuss the corrections to the membrane elastic constants given by
Eqs. (\ref{deltasigma})-(\ref{deltakm}). In particular
we obtain the dependance of these elastic moduli on the ionic strength of the electrolyte,
and on the ion conductivity of the membrane $G$.
Fig. \ref{DSdetailG0}a) displays separately the contributions to the
surface tension as a function of the inverse Debye length
$\kappa$. The dashed line represents the contribution from the
Debye layers, the dash-dotted line represents the contribution
from the field inside the membrane and the solid line is the sum
of both contributions. The value of the inverse Debye length
$\kappa$, varies from $\kappa\simeq10^6{\rm m}^{-1}$
($\lambda_D=1\mu{\rm m}$) for pure water to
$\kappa=3.3\cdot10^9{\rm m}^{-1}$ ($\lambda_D\simeq0.3{\rm nm}$)
for 1M NaCl \cite{andelman}. Fig. \ref{DSdetailG0}b) shows the
respective contributions to the membrane bending modulus. In this
figure we have assumed that the membrane is non-conductive
($G=0$).
As shown in Fig. \ref{DSdetailG0}, the contributions from the
Debye layers dominate for low salt ($\kappa<5\cdot10^6{\rm m}^{-1}$ for $\Delta\Sigma$
and $\kappa<10^8{\rm m}^{-1}$ for $\Delta K$). For high salt, the contributions
from the membrane dominate, and approach the limiting values discussed above.
In the case of zero conductivity, both the Debye and the inside contribution
to the surface tension are always negative, and there is
good agreement with calculations for non-conductive membranes \cite{lomholt_elect,lacosteEPJE}.
Fig. \ref{DSdetailGfin} displays the corrections to the elastic coefficients in
the case where a finite membrane current $j_m$ is present, induced by a small
conductivity $G=0.1 S{\rm m}^{-2}$, for otherwise
unchanged parameters.
We find that this rather small conductivity has already a large effect on both moduli:
first, the effect of the Debye layers on the surface tension is suppressed
and the contribution from the membrane is dominating. Second, the overall contribution
gets relevant for higher salt than in the non-conductive case. The effect on the bending modulus
is even more significant: although the Debye contribution is still dominating for
about $\kappa<10^8{\rm m}^{-1}$, it is much smaller in amplitude than in the
non-conductive case (around $10^{-20}J$ compared to $10^{-17}J$ at $\kappa\simeq5\cdot10^6{\rm m}^{-1}$)
and furthermore becomes non-monotonous \cite{lacosteEPJE}.

\begin{figure}[t]
    \centering
    \includegraphics[width=0.45\textwidth]{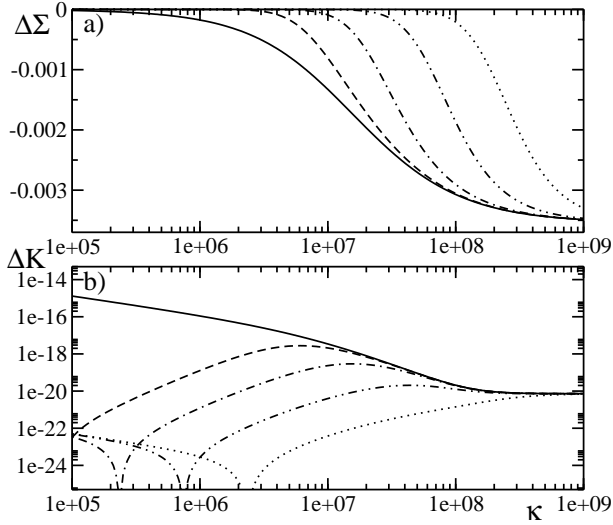}
    \caption{\label{DSvarG}
    Electrostatic corrections
    to the surface tension (panel a) and to the bending modulus (panel b)
    as a function of $\kappa$ for different membrane conductivities.
    Parameters are as in the two previous figures except for $G$:
    solid line $G=0$; dashed line $G=0.01$; dash-dotted line $G=0.1$; dash-two-dots line $G=1$;
    dotted line $G=10$ in units of $S{\rm m}^{-2}$.
    }
\end{figure}

The effect of membrane conductivity is highlighted in Fig. \ref{DSvarG},
where the total contributions to surface tension (panel a) and
bending modulus (panel b) are shown as a function of the inverse
Debye length $\kappa$ for conductivities in the range
$G=0.01-10S{\rm m}^{-2}$, and otherwise unchanged parameters.
We find that, except for the high salt limit, the bending modulus correction tends to be
reduced by increasing membrane conductivity.
To give some numbers, for a non-conductive membrane ($G=0$) and $V=1{\rm V}$,
the jump in the charge density for $\kappa=2\cdot10^7{\rm m}^{-1}$
is $\rho_m=1.5\cdot10^{23}\frac{e}{\rm{m}^3}$.
Already a small value of the conductivity as $G=0.1 S{\rm m}^{-2}$
halves the charge density to
$\rho_m=8.6\cdot10^{22}\frac{e}{\rm{m}^3}$ and creates the
current density $j_m=3.7\cdot10^{17}\frac{e}{\rm{m}^2 \rm{s}}$,
or $60\cdot10^{-3} \frac{{\rm A}}{{\rm m}^2}$.
Conductivities of biological membranes
can be as high as
$G=10 S{\rm m}^{-2}$, which is the value for a squid axon,
corresponding to a density of potassium channels of 0.5$\mu$m$^{-2}$ \cite{hille}.

We also mention that the distance between the electrodes $L$, i.e.
the confinement, is a relevant parameter and influences the shape
of Figs. \ref{DSdetailG0} and \ref{DSdetailGfin}. Here we have
used a macroscopic distance ($L=1{\rm mm}$), corresponding to the
experiments mentioned above. If $L$ was instead of the order of
microns
the suppression of the bending modulus correction due to
conductivity would be much less pronounced and the corresponding
figure would become similar to the one given in Ref.
\cite{lacosteEPJE}. Moreover, for high enough membrane
conductivity, the Debye layer contribution to the surface tension
can become positive, i.e. stabilizing. In fact, the sign of
$\Delta\Sigma_\kappa$ is governed by a factor $-\frac{\epsilon
D\kappa^2-2G\dm}{\left(\epsilon D \kappa^2+2GL\right)^2}$. In a
way similar as discussed in section \ref{base}
concerning the sign of $\rho_m$,
for $G>\frac{\epsilon D \kappa^2}{2\dm}$ there is a sign change, rendering the correction
positive for small $\kappa$. However, the denominator containing
the distance $L$ between the electrodes suppresses this effect for
macroscopic distances. It can be seen only if $L$ is small, e.g.
$L=1\mu{\rm m}$ as used in Ref. \cite{lacosteEPJE}.
We note that the micron scale is particularly relevant to experiments with
cell membranes submitted to electric fields \cite{sachs}. It is also relevant
to experiments that one could propose to test these ideas using
microfluidics devices.

\subsection{Membrane instability}
\label{membr_instab}

\begin{figure}[t]
    \centering
    \includegraphics[width=0.45\textwidth]{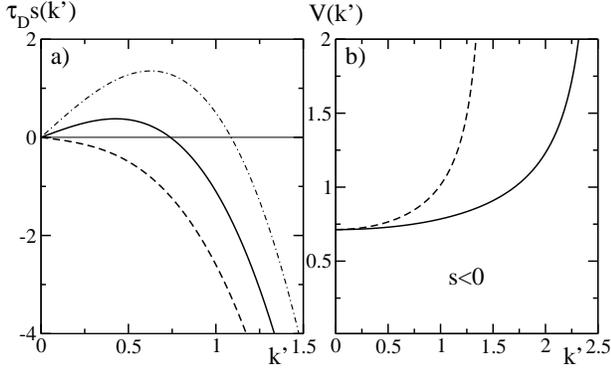}
    \caption{\label{dispfig}
    a) The renormalized growth rate or dispersion relation, $\tau_D s$,
    as a function of the rescaled wave number $k'=\kpm/\kappa$ for three voltages:
    $V=0.7{\rm V}$ (dashed line),
    $V=0.75{\rm V}$ (solid line), $V=0.8{\rm V}$ (dash-dotted line).
    b) The neutral curve (solid line) separating the regions of $s>0$ and $s<0$,
    and the fastest growing wave number $k_{fg}$ (dashed line) in the plane
    voltage vs. rescaled wave number $k'=\kpm/\kappa$.
    Parameters as previously except: no conductivity, $G=0$; $\kappa=2\cdot 10^{7}{\rm m}^{-1}$;
    $\Sigma_0=1$mNm$^{-1}$; $K_0=10k_B T$.
    }
\end{figure}

Since the corrections to the membrane surface tension are
typically negative (with the exception mentioned above), they can
overcome the bare surface tension $\Sigma_0$. At this point, an
instability towards membrane undulations sets in
\cite{pierre}. Our theory is able to go beyond previous modeling
of this instability (still for early stages of the instability),
which were limited to the high salt limit and
did not include electrostatic corrections to the bending modulus
or hydrodynamic effects associated with the modulus $\Gamma$. The
linear growth rate of the membrane fluctuations is given by Eq.
(\ref{disp_fin}) and has the form
\beq\label{growthsimple}
\eta s(k)=-\frac{1}{4}\Sigma_{eff} k-\Gamma_\kappa
k^2-\frac{1}{4}K_{eff}k^3\,,
\eeq
where we have written simply $k$
for $\kpm$ and introduced the effective surface tension and
modulus, $\Sigma_{eff}=\Sigma_0+\Delta\Sigma$, $K_{eff}=K_0+\Delta
K$. Fig. \ref{dispfig}a) shows this growth rate, or dispersion
relation, in rescaled units where we scaled the wave vector by
$\kappa$, $k'=k/\kappa$, and the time by the typical time for
ions to diffuse a Debye length, $\tau_D=\frac{1}{D\kappa^2}$.
The parameters are the same as in the previous sections for a non-conductive membrane, i.e. $G=0$.
The control parameter is the external voltage $V$.
Fig. \ref{dispfig}a) shows the growth rate for three different levels of the voltage:
the dashed line is for $V=0.7{\rm V}$, which lies below the threshold of the instability,
all wave numbers are damped and the membrane is stable.
The solid and the dash-dotted line correspond to $V=0.75{\rm V}$ and $V=0.8{\rm V}$ and are above threshold.
A certain window of wave numbers $k\in ]0,k_{max}(V)[$ has positive growth rates and the
membrane is thus unstable. This window gets larger with increasing voltage. The linear growth will
be dominated by the maximum of the growth rate defining the fastest growing wave number
$k_{fg}$. Given Eq. (\ref{growthsimple}), one easily calculates
\beq
k_{fg}&=&\frac{4/3}{K_{eff}}\left(-\Gamma_\kappa+\sqrt{\Gamma_\kappa^2-\frac{3}{16}K_{eff}\Sigma_{eff}}\right)\,,\quad%\,\,\\
%k_{max}&=&\frac{2}{K_{eff}}\left(-\Gamma_\kappa+\sqrt{\Gamma_\kappa^2-\frac{1}{4}K_{eff}\Sigma_{eff}}\right)\,,
\eeq
for $\Sigma_{eff}<0$.

The same information given by the dispersion relation can be expressed 
by the so-called neutral curve which is
shown in Fig. \ref{dispfig}b). This curve, given by the solid line, separates
the negative (below) from the positive (above) growth rates in the control parameter-wave number
plane. If the voltage is below the section of the neutral curve with the voltage-axis,
the system is stable. Otherwise a certain band
of wave numbers is unstable.
The position of the fastest growing mode $k_{fg}$ is given by the dashed line.

Since we have the dispersion relation in analytical form, in principle
one has formulas for all relevant observables like the threshold voltage $V_c$.
In terms of the system parameters, however, they are quite lengthy.
The threshold voltage is given by the change of sign of the leading order contribution in $s(\kpm)$.
In the non-conductive case it has the simple form
\beq
V_c^2(G=0)=\frac{\Sigma_0 d(2+\kappa\dm)^2}{\kappa(\kappa\epsilon_m\dm^2+\epsilon d)}\,.
\eeq
Since both $\rho_m$ (and $j_m$ in case of $G\neq0$) are proportional to the voltage,
as expected the critical voltage scales like $V_c\propto\sqrt{\Sigma_0}$.
In the limit of small membrane conductivity, one gets to leading order
(using that $L$ is macroscopic)
\beq
V_c^2=V_c^2(G=0)\left(1+\frac{4GL}{\epsilon D\kappa^2}\right)\,.
\eeq
Thus membrane conductivity increases the voltage value needed to cross
the instability.
In the limit of high salt, $\kappa\rightarrow\infty$, one regains the known result
$V_c^2=\Sigma_0 d/\epsilon_m$.
The typical wavelength of the membrane undulations above threshold (i.e. the one
of the fastest growing mode) for parameters as in Fig. \ref{dispfig}
is of order $\lambda=\frac{2\pi}{0.5\kappa}\simeq0.25\mu{\rm m}\simeq 12.5\lambda_D$,
so several times the Debye length.

\begin{figure}[t!]
\begin{center}
    \includegraphics[width=.44\textwidth]{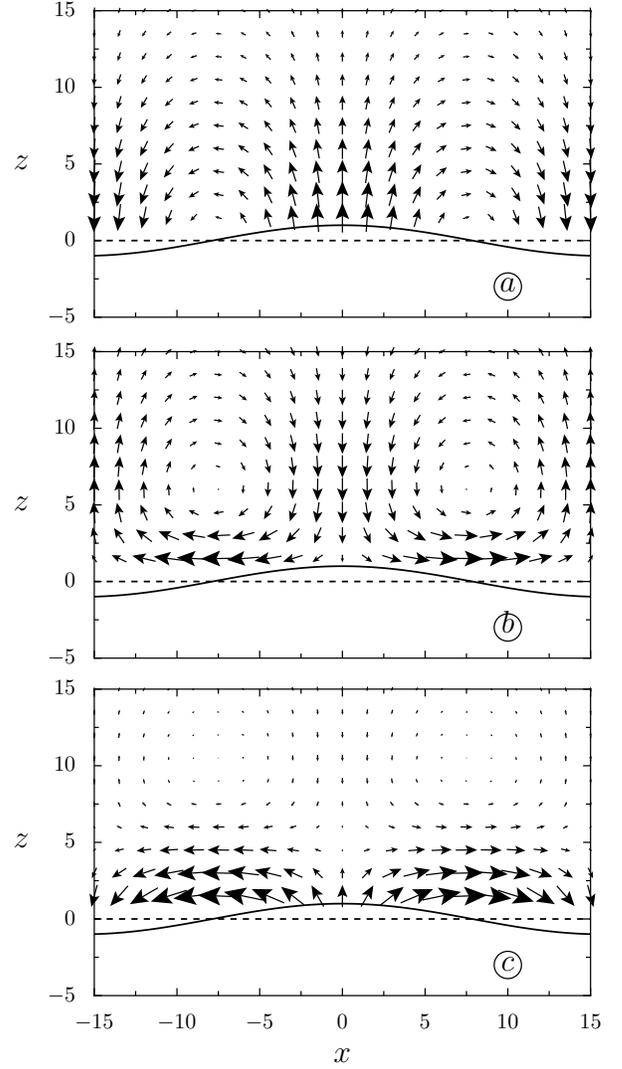}
    \caption{\label{flowfig} Representation of the flows around the membrane
    beyond the instability threshold. The orientation
    of the electric field is towards negative values of $z$.
    Panel a) shows the flow generated by the membrane bending instability
    (terms proportional to $s$ in Eqs. (\ref{vzfull}, \ref{vpfull})).
    Panel b) shows the ICEO flow (terms proportional to $\al$ in Eqs. (\ref{vzfull}, \ref{vpfull})).
    Finally, panel c) shows the actual flow, %as given by Eqs. (\ref{vzfull}, \ref{vpfull}),
    which is the superposition of the former two
    and results in a strong flow near the membrane, oriented
    parallel to the surface. Both axes are scaled by the Debye length $\kappa^{-1}$.
    Parameters are as in previous figures except $V=3.165{\rm V}$,
    $\kappa=10^{7}{\rm m}^{-1}$, $G=10 S{\rm m}^{-2}$ and $L=10\mu{\rm m}$.
    }
\end{center}
\end{figure}

\subsection{ICEO flows}

We now discuss the form of the fluid flows which arise near the
membrane when it is driven by ionic currents. Fig. \ref{flowfig}c)
shows the flow field for a high membrane conductivity and low
salt, in the regime where the membrane is unstable due to the
electrostatic correction to the surface tension and thus starts to
undulate. This figure was generated by selecting the fastest
growing wave number $k_{fg}$, defined in section
\ref{membr_instab}, and using the respective maximum growth rate
$s(k_{fg})$. Since this wave number has the fastest growth
rate in the linear regime, it will dominate the initial behavior.
The shape of the membrane undulation is represented as the black solid curves
in all plots of Fig. \ref{flowfig}. The resulting flow, shown in
Fig. \ref{flowfig}c) is a superposition of two distinct flows:
first, the typical flow associated to a membrane bending mode
\cite{BrochardLennon,levine} as shown in Fig. \ref{flowfig}a).
This contribution corresponds to the terms proportional to the growth
rate $s$ in Eqs. (\ref{vzfull}, \ref{vpfull}). Second, the flow
associated with the remaining terms in Eqs.
(\ref{vzfull}, \ref{vpfull}), proportional to $\al$. This
contribution yields the typical counter-rotating vortices of an
ICEO flow \cite{bazant:2004}, as shown in Fig. \ref{flowfig}b).
Clearly, the superposition of these two flow contributions, as
shown in \ref{flowfig}c), results in a parallel flow close to the
membrane, in contrast to the usual bending mode flow given by Fig.
\ref{flowfig}a).

Since the jump of the charge density $\rho_m>0$
for biological membranes (and $j_m\geq 0$ by
definition), the induced flow occurs for this case in the same
direction as in standard ICEO flows. Note that
an inverse ICEO flow was obtained in
\cite{lacosteEPJE}, due to the opposite sign of
$\rho_m$ obtained with the simple but unrealistic BC
Eq.~(\ref{BC_simple}).
Also, the situation of ICEO flows is less general as suggested earlier:
For most parameters (modest conductivities, not too low salt)
the flow generated by membrane bending is usually dominating
and hides the small ICEO contribution. This is due to the fact that
the former is proportional to $s$ which has contributions $\propto\rho_m^2$,
while the ICEO flow is $\propto\rho_m j_m\ll\rho_m^2$.
Thus, to see the situation given by Fig. \ref{flowfig},
a high membrane conductivity $G$ is needed. Second, one needs low salt, since
otherwise the membrane instability is shifted to very high voltages.
Also, since for macroscopic electrode distances $L$ (of order millimeter) and high
conductivity the voltage needed to induce the instability is very high,
we have used a microscopic electrode distance $L=10\mu{\rm m}$.
While it might still be possible to see these flows for higher
salt and macroscopic electrode separations, such situations will
be clearly far beyond the Debye-H\"uckel approximation used so
far.

The ICEO flows near the membrane
could also become relevant once the system has reached a steady-state.
Indeed in the case of lipid vesicles for instance, non-linear effects associated
with the conservation of the number of lipids on the vesicle \cite{pierre}
guarantee a saturation of the membrane fluctuations (for not too high voltages that might
lead to vesicle rupture), as compared to the case of the
planar membrane considered here. Since the membrane fluctuations are
confined by non-linear effects and become quasi-stationary in the long time limit,
the system can reach a well defined non-equilibrium steady state.
In this non-equilibrium steady state it might well be possible that
fluid flows still arise due to ionic currents going through the membrane,
while the initial flow associated with the membrane bending mode has disappeared.

%%%%%%%%%%%%%%%%%%%%%%%%%%%%%%%%%%%%%%%%%%%%%%%%%%%%%%%%%%%%%%%%%%%%%%%%%%%%%%%
\section{Conclusions and perspective}
\label{Concl}
%%%%%%%%%%%%%%%%%%%%%%%%%%%%%%%%%%%%%%%%%%%%%%%%%%%%%%%%%%%%%%%%%%%%%%%%%%%%%%%

This paper offers a route to describe capacitive effects near a
conductive lipid membrane while keeping the simplicity of the zero thickness
approximation on which most of the literature on lipid membranes
is based. These capacitive effects are the main player in the
corrections to the elastic moduli of membranes driven by an
electric field or by internal pumps or ionic channels. The present
theory goes beyond available descriptions by
including non-equilibrium effects which arise due to ionic
membrane currents. These ionic currents have a similar form as the ICEO
flows studied in the context of microfluidics and can modify the fluid flows
around the membrane from usual bending dominated flow towards flow concentration
close to and parallel to the membrane.

Our approach is sufficiently simple to be the starting point for further
generalizations, which could include various nonlinear effects:
non-linear elastic terms associated with the membrane or the cytoskeleton
in case of a biological membrane, non-linear current-voltage response of the channels.
Also density fluctuations of the ion channels and the experimentally simpler case
of an AC electric field should be investigated.
Further theoretical work is also needed to extend the model to higher
voltage where the Debye-H\"uckel approximation breaks down.
In fact, this approximation
%Further work is also needed on the theory side. One limitation of
%the present model is the Debye-H\"uckel
%approximation, which
only holds when the potential satisfies
everywhere the condition $\frac{e\phi}{4k_B T}\ll1$ \cite{hunter}.
For the experiments on supported membranes discussed above,
one finds that the charge accumulation on the membrane
is too large for this approximation to hold, since one has
$\frac{e\phi}{4k_B T} \simeq 2$. It would thus be relevant
to solve the nonlinear Poisson-Nernst-Planck equations and
otherwise proceed similarly as proposed in this work,
in order to describe the behavior of membranes surrounded by high charge densities.
%This
%extension is required in the case of the experiments on supported
%lipid membranes, since very large electric fields have been used
%in that experiment.
Furthermore, it would be interesting to
investigate a model suitable for small system sizes,
since much of the results of this paper are based on the
assumption that the system size $L$ is much larger than all other
length scales in the problem, as well as for more realistic
boundary conditions at the electrodes.

We would like to thank Thierry Charitat and Pierre Sens
for fruitful discussions and Luis Dinis for a careful reading
of the manuscript.
F.Z. acknowledges financial support from the German Science
Foundation (DFG), M.Z.B. support from the U.S.
National Science Foundation under Contract DMS-0707641 and 
D.L. from the Indo-French Center for the Promotion
of Advanced Research under Grant No. 3502 and ANR for funding.

\appendix

%%%%%%%%%%%%%%%%%%%%%%%%%%%%%%%%%%%%%%%%%%%%%%%%%%%%%%%%%%%%%%%%%%%%%%%%%%%%%%%%%%%
\section{Contribution of the internal field to the Maxwell stresses}
\label{ins_memb_new}
%%%%%%%%%%%%%%%%%%%%%%%%%%%%%%%%%%%%%%%%%%%%%%%%%%%%%%%%%%%%%%%%%%%%%%%%%%%%%%%%%%%

Here we are interested in the contribution of the internal field
inside the membrane to the stress, which is
\beq
[\tau^m_{zz,1}]_{z=0}
=\epsilon_m\left[(\p_z\phi^m_0)(\p_z\phi^m_1)\right]_{z=\pm d/2}\,.
\eeq
 Since the internal field at zeroth order is constant inside the membrane
due to the symmetry of the problem,
this expression simplifies into
\beq
[\tau^m_{zz,1}]_{z=0}=-2\epsm E^m_0\left(\p_z\phi^m_1\right)_{|z=+d/2}\,.
\eeq
The first-order field in the membrane is given by (use the symmetry or
cf. Ref. \cite{lacosteEPJE} for details)
\beq
\phi^m_1(\kpm,z)=\phi^m_1(\kpm,d/2)\frac{e^{\kpm d/2}}{e^{\kpm d}+1}\left(e^{\kpm z}+e^{-\kpm z}\right),\,\,\,
\eeq
which leads us to
\beq\label{tau_m1}
[\tau^m_{zz,1}]_{z=0}=-2\epsm E^m_0\phi^m_1\left(\kpm,\frac{d}{2}\right)\kpm
\frac{e^{\kpm d}-1}{e^{\kpm d}+1}\,.
\eeq

To obtain an expression for $\phi^m_1\left(\kpm,\frac{d}{2}\right)$,
to linear order in $h$, one can write \cite{lacosteEPJE}
\beq
\phi^m_1\hspace{-1mm}\left(\kpm,\frac{d}{2}\right)=\phi_{1}\hspace{-1mm}\left(\kpm,\frac{d}{2}\right)
-h\left(\p_z\phi^m_0-\p_z\phi_{0}\right)_{|z=d/2}\,.\quad
\eeq
For the outside potential at the membrane
we can approximately use
$\phi_1(\kpm,z)=-\frac{2}{\epsilon}\rho_m\frac{h(\kpm)}{l}$
here (since the exponential decay like $e^{-lz}$ starts at the membrane).
After expansion in $\kpm$
(and assuming $L\gg d$)
we get
\beq
\phi^m_1\left(\kpm,\frac{d}{2}\right)
&=&h(\kpm)\left[E^m_0+\frac{\rho_m}{\eps\kappa^3}\kpm^2\right]\,.
\eeq
Using Eq. (\ref{tau_m1}), for the stress difference, $[\tau^m_{zz,1}]_{z=0}$,
up to order $\kpm^4$ this exactly yields Eq.~(\ref{contrib_ins}) given above.

%%%%%%%%%%%%%%%%%%%%%%%%%%%%%%%%%%%%%%%%%%%%%%%%%%%%%%%%%%%%%%%%%%%%%%%%%%%%%%%
\section{Electrostatic contribution to the surface tension from the stress tensor}
\label{Widom}
%%%%%%%%%%%%%%%%%%%%%%%%%%%%%%%%%%%%%%%%%%%%%%%%%%%%%%%%%%%%%%%%%%%%%%%%%%%%%%%

In this appendix, we give an alternative approach for the derivation of
the membrane tension, which avoids solving for the
fluid flow around the membrane. Here
the tension is expressed as an integral over the lateral pressure
profile deviation or, equivalently, over the excess lateral stress \cite{widom}.

Let us call $S$ a closed surface englobing the membrane
with the normal vector field ${\bf n}$. We choose ${\bf x}$
to represent the direction of the lateral stress. The force acting
on the surface $S$ in the ${\bf x}$-direction  can be calculated
from the stress tensor defined in Eq.~(\ref{stresstensor}) as
\beq
\label{def_F_x}
F_x=\int_S {\bf x} \cdot \tau \cdot {\bf n} \,\,dS.
\eeq
Since $\tau$ is divergence free, the surface $S$ can be
deformed into an arbitrary other surface englobing $S$, for convenience to
a cube of size $L$.
It is easy to see that the integral in Eq.~(\ref{def_F_x})
is non-zero only on the faces of the cube with the
normal along $\pm{\bf x}$. With $dS=L dz$ and $\Delta
\Sigma=F_x/L$ on the face where ${\bf n}=+{\bf x}$, we arrive at \cite{lomholt_elect}
\beq \label{new delta sigma}
\Delta \Sigma=\int_{-L/2}^{L/2} \tau_{xx}(z) dz.
\eeq

Eq.~(\ref{stresstensor}) implies
$\tau_{xx}(z)=-P_0(z)-\frac{\epsilon}{2} (\partial_z \phi_0)^2$,
where $\phi_0(z)$ is the
potential and $P_0(z)$
the pressure profile in the base state. The resulting tension
$\Delta \Sigma$ is identical to Eqs.~(41-43) of Ref.~\cite{lacosteEPJE},
where it was expressed as a sum of two
terms, $\Sigma_0$ and $\Sigma_1$. A cancellation of the dependance on
$L$ occurred
in the sum of these two terms, as it should be. The
present derivation fully justifies this point since the choice of
the deformed surface was arbitrary by construction.
Using Eqs.~(\ref{field}, \ref{pressure profile base state}, \ref{new delta sigma}), we obtain
\beq
\Delta\Sigma_\kappa=-4\frac{\rho_m^2}{\epsilon\kappa^3}-16\frac{\rho_m
j_m}{\epsilon \kappa^4 D},
\eeq which is exactly the result for the
contribution of the Debye layers to the tension using the
hydrodynamic approach, Eq.~(\ref{deltasigma}). A similar
calculation gives contribution to the tension from inside the membrane,
which reads 
\beq
\Delta\Sigma_m=-\epsilon_m d (E_m^0)^2.
\eeq
Note that the
same method could be used to calculate the change of spontaneous
curvature induced by an electric potential in the asymmetric case
\cite{lomholt_elect}.

%\bibliography{biophys,mes_publis}

\end{document}